\documentclass[prd,superscriptaddress,twocolumn,nofootinbib]{revtex4-2}
\usepackage{color}
\usepackage{bm}
\usepackage{graphicx}  
\usepackage{dcolumn}   
\usepackage{bm}        
\usepackage[english]{babel}
\usepackage{booktabs,multirow,array}
\usepackage[bottom]{footmisc}
\usepackage{dblfnote}
\usepackage{mathtools}
\usepackage{amsmath}
\usepackage{physics}
\DFNalwaysdouble 
\usepackage[colorlinks]{hyperref}
\definecolor{LinkColor}{rgb}{0.75, 0, 0}
\definecolor{CiteColor}{rgb}{0, 0.5, 0.5}
\definecolor{UrlColor}{rgb}{0, 0, 0.75}
\hypersetup{linkcolor=LinkColor}
\hypersetup{citecolor=CiteColor}
\hypersetup{urlcolor=UrlColor}

\begin{document}

\title{Thermal radiation in curved spacetime using influence functional formalism }

\author{Chiranjeeb Singha}
\email{chiranjeeb.singha@saha.ac.in}
\affiliation{ Theory Division,
Saha Institute of Nuclear Physics,
Kolkata 700064,
India }

\author{Subhashish Banerjee}
\email{subhashish@iitj.ac.in}
\affiliation{Indian Institute of Technology Jodhpur, Jodhpur 342011, India}


\date{\today}

\begin{abstract}

Generalizing to relativistic
exponential scaling and using the theory of noise from
quantum fluctuations, it has been shown that one vacuum (Rindler, Hartle-Hawking,
or Gibbons-Hawking for the cases of the uniformly accelerated detector, black hole, and de-Sitter universe, respectively) can be understood as resulting from the scaling of quantum noise in another vacuum.
We explore this idea more generally to establish a flat spacetime and curved spacetime analogy. For this purpose, we start by examining noise kernels for free fields in some well-known curved spacetimes, e.g., the spacetime of a charged black hole, the spacetime of a Kerr black hole, Schwarzschild-de Sitter, Schwarzschild anti-de Sitter, and Reissner-Nordstrom de-Sitter spacetimes. Here, we consider a maximal analytical extension for all these spacetimes and different vacuum states. We show that the exponential scale transformation is responsible for the thermal nature of radiation.  
\end{abstract}
\maketitle
 \section{introduction}
 
The Unruh effect refers to the thermal fluctuations experienced by a detector while undergoing linear motion with uniform acceleration in a Minkowski vacuum \cite{Unruh, DEWITT1975295, Davies_1975, PhysRevD.7.2850, Omkar:2014hba, Banerjee:2016ahn}. This
thermality can be demonstrated by tracing the vacuum state of the field over the modes beyond the accelerated detector's event horizon. However, the event horizon is well-defined only if
the detector moves with eternal uniform linear acceleration, wherein the particle's speed
asymptotically approaches the speed of light, entailing the formation of an event horizon. In
contrast, in the circular case, velocity changes direction, but its magnitude remains constant, and there is no event horizon. In the realistic Unruh case, eternal uniform linear acceleration is
impossible; hence the notion of the horizon is difficult to envisage. In \cite{Doukas:2013noa} the effect was studied as
a kinematic effect in terms of influence functionals.
In this context, also see \cite{Sinha:2002iq}. Thermal radiance from a black hole or observed by an accelerated detector is usually viewed as a geometric effect related to the existence of an event horizon. 
In
\cite{Hu:1996vu, PhysRevD.49.6612}, it was proposed that the detection of thermal radiance in these systems is a local, kinematic
effect arising from the vacuum being subjected to a relativistic exponential scale transformation.\\

Generalizing to relativistic exponential scaling and using the theory of noise kernel from quantum
fluctuations \cite{Candelas:1977}, it has been shown that one vacuum (Rindler, Hartle-Hawking,
or Gibbons-Hawking for the cases of the uniformly accelerated detector, black hole, and de Sitter
universe, respectively) can be understood as resulting from the scaling of quantum noise in
another.\\

This paper explores the idea of relativistic exponential scaling and influence function formalism more generally to establish a flat spacetime and curved spacetime analogy.
For this purpose, in sec. \ref{relativistic scaling}, we briefly discuss relativistic exponential scaling. 
In sec. \ref{influence function}, a sketch of the influence functional formalism is provided.
The derivation of this formalism in different spacetimes is essentially the same as in the Unruh effect.
Hence, in sec. \ref{unruh effect}, the key derivation of the Unruh effect using exponential scale transformation and influence functional formalism is brought out. This formalism is applied in the other examples discussed. Thus, for example, the Hawking radiation for 1+1 dimensional charged black hole, Kerr black hole, 1+1 dimensional Schwarzschild de-Sitter, Schwarzschild anti-de Sitter, and Reissner-Nordstrom de-Sitter spacetimes are studied in secs. \ref{charged black hole}, \ref{kerr black hole}, \ref{Schwarschild de-Sitter spacetime}, \ref{Schwarschild anti-de Sitter spacetime}, \ref{Reissner-Nordstrom de-Sitter} respectively, using exponential scale transformation and the influence functional formalism. Used is made of maximal analytical extension for all these spacetimes and different vacuum states. The temperature obtained is consistent with that in the literature, obtained in different contexts. We finally make our conclusions. 

\section{relativistic exponential scaling} \label{relativistic scaling}

There exist several well-established methods in quantum field theory in curved spacetimes \citep{book:75900, BELL1983} to study the particle creation from a black hole or Unruh effect. The central aspect of the Unruh effect is that the vacuum state where
thermal radiance is observed (\textit{e.g.,} the Rindler vacuum), is related to the inertial vacuum state (Minkowski vacuum) by an exponential
scale transformation. Similarly, for the Hawking radiation, the vacuum state where
thermal radiance is observed (\textit{e.g.,} the Schwarzschild vacuum which is also known as the Boulware vacuum), is related to the inertial vacuum state (\textit{e.g.,} Kruskal vacuum which is also known as the Hartle-Hawking vacuum) by an exponential
scale transformation.
Due to the presence of the exponential terms, 
Rindler spacetime used by accelerated observers covers only a
wedge-shaped region of Minkowski spacetime. Similarly, for the Hawking effect, the Schwarzschild coordinate covers only a wedge-shaped patch of the Kruskal coordinate \cite{Hu:1996vu, PhysRevD.49.6612}. Hence, using these coordinate systems,
one can see the Hawking radiation and the Unruh effect \cite{hawking_black_1974, Unruh}. One can also consider another vacuum state such as the Unruh state for calculating Hawking radiation \cite{book:75900, Singleton:2011vh}. The Unruh vacuum state corresponds to the Boulware vacuum in the far past and the Hartle-Hawking vacuum in the far future \cite{Singleton:2011vh, Juarez-Aubry:2021tae}. In this paper, we also show that the exponential scale transformation is responsible for the thermal nature of radiation of these spacetimes, thus highlighting a uniform approach to this phenomenon. Further, this also strives to build up a geometry, quantum statistical mechanical equivalence.

\section{INFLUENCE FUNCTIONAL}\label{influence function}

We consider the scenario wherein a detector is used to probe the unperturbed state of a scalar field. The detector and the massless scalar field can be considered to be the system and its bath, respectively. The system interacts weakly with the massless scalar field. We assume that at a given initial time, the system and the environment are uncorrelated. Thus, the total density matrix of the total system (system+environment) at the initial time can be written as the density matrix of the system outer product with the density matrix of the environment. As we are interested in the time evolution of the system of interest, which, here, is the detector, the field degrees of freedom are traced out. This provides the reduced density matrix of the detector, taking into account the influence of the field. This is ensconced in the form of the influence functional  \cite{FEYNMAN1963118, CALDEIRA1983587,GRABERT1988115, PhysRevD.45.2843, book:408298, sbbook, sbQBM}, which can be generically shown to be

\begin{widetext}
\begin{eqnarray}
\mathcal{F} [x,x']=\exp\left\{ -\frac{i}{\hbar} \int_{0}^{\tau} d\tau \int_{0}^{s} d\tau' (x(\tau) - x(\tau'))\left[\mu(\tau,\tau')(x(\tau') + x'(\tau'))-i \nu (\tau,\tau')(x(\tau') - x'(\tau'))\right]  \right\},
\end{eqnarray}
\end{widetext}
where $\nu(t,t')$ and $\mu(t,t')$ are the noise and dissipation kernels, respectively. These kernels can be written compactly as 
\begin{eqnarray}
\varsigma (t(\tau), t(\tau'))&=&\int dk~I(k,\tau,\tau')~ e^{-i \omega[t(\tau)-t(\tau')]}\nonumber\\
&=&\nu(t,t')+i \mu(t,t')~,
\end{eqnarray}
where $I(k,\tau,\tau')$  is the field (bath) spectral density.\\

The influence functional is a path integral approach to study the dynamics of the system of interest taking into account the effects of the environment, sometimes called the reservoir. From the influence functional, characterizing the environment, one obtains
what is known as the propagator. The propagator can then be used as a capsule to generate the final state of the system of interest, given its initial state, hence the name propagator. This provides a flexible tool to approach a vast variety of problems, ranging from the early universe, black holes \cite{hpz93} to decoherence in the quantum to classical transition \cite{hartle}. Here, a paradigm model is that of quantum Brownian motion \cite{CALDEIRA1983587,GRABERT1988115}. The noise and dissipation kernels characterizing the influence kernel are determined by the spectral density \cite{weiss}. In order to obtain true irreversible dynamics,  a continuous distribution of bath modes can be introduced, such that the spectral density is represented by a smooth function of the environment's frequencies.

In this work, we explore the idea of relativistic exponential scaling and influence function formalism more generally with the aim to establish a flat spacetime and curved spacetime analogy. For this purpose, we  examine noise kernels for free fields in some well-known curved spacetimes, e.g., the spacetimes of a charged black hole, a Kerr black hole, Schwarzschild-de Sitter, Schwarzschild anti-de Sitter, and Reissner-Nordstrom de-Sitter spacetimes. The essential derivation of this formalism is the one involved in the Unruh effect because the relation between the tortoise and the Kruskal coordinates is similar to the  Rindler coordinate and the inertial coordinate. We briefly review the key derivation of the Unruh effect using exponential scale transformation and influence functional formalism in the next section.

\section{Unruh effect } \label{unruh effect}

A uniformly accelerated observer can see a part of Minkowski spacetime. The line element for this observer in Minkowski spacetime is given by \cite{Crispino:2007eb, Unruh, PhysRevD.7.2850},
\begin{eqnarray}\label{rindler}
d s^2  &=& e^{2 a \xi}(-d \tau^2 +d \xi^2)+dy^2+dz^2~. 
\end{eqnarray} 
The observer is uniformly accelerated in $x$ direction, and $a$ is the four acceleration vector. On the other  hand, the Minkowski metric is given by,
\begin{equation}\label{Minkowski1}
 d{s}^2=-d t^2+d x^2+d y^2+d z^2~.
\end{equation}
The $\{y,z\}$ coordinates are the same here. The following transformation connects the accelerated observer and the inertial observer,
\begin{eqnarray}
t (\tau) = \frac{1}{a}e^{a \xi} \sinh (a \tau), \quad x (\tau) = \frac{1}{a}e^{a \xi} \cosh(a \tau)~.
\end{eqnarray}
Here, we see that the Rindler coordinates are related to the inertial coordinates by an exponential scale transformation. Hence, Rindler spacetime used by accelerated observers covers only a wedge-shaped region of Minkowski spacetime. Now we consider a two-dimensional massless scalar field $\phi$ in flat spacetime with mode decomposition
\begin{equation}
\phi (x)=\sqrt{2/L}\sum_k\left[q^+_k \cos kx+q^-_k \sin kx\right]~.
\end{equation}
The Lagrangian for the field can be expressed as a sum of oscillators with amplitude $q^{\pm}_k$ for each mode as
\begin{equation}
L(s)=\frac{1}{2}\sum^{+ -}_\sigma\sum_k\left[\left(\dot{q^\sigma_k}\right)^2-k^2 \left(q^\sigma_k\right)^2\right]~.
\end{equation}
Here we consider an observer undergoing constant acceleration $a$ in this field with the following trajectory
\begin{equation}
x(\tau)=\frac{1}{a} \cosh(a \tau)~~,~~t(\tau)=\frac{1}{a} \sinh(a \tau)~.\label{rindleraccl}
\end{equation}
We want to show via the influence functional method that the observer detects thermal radiation. The system-field interaction is taken as \cite{PhysRevD.47.4525, PhysRevD.49.6612}, 
\begin{equation}
\mathcal{L}_{int}(x)=- \epsilon r \phi(x)\delta(x(\tau))~.
\end{equation}
They are coupled at the spatial point $x(\tau)$ with coupling strength $\epsilon$ and $r$ is the detector's internal coordinate. Integrating out the spatial variable, we find that 
\begin{equation}
\mathcal{L}_{int}(\tau)=\int \mathcal{L}_{int}(x) dx =- \epsilon r \phi(x(\tau))~.
\end{equation}
The scalar field is characterized by the spectral density, which in the discrete case is given by \cite{PhysRevD.47.4525, PhysRevD.49.6612}, 
\begin{equation}
I(k,\tau,\tau')= \sum^{+ -}_\sigma\sum_n\frac{\delta(k-k_n)~c^\sigma_n(\tau)~c^\sigma_n(\tau')}{2 \omega_n}~.
\end{equation}
Here $k_n=\omega_n=|k_n|$, as we take a massless scalar field. $ c^+_n (\tau)= \epsilon\sqrt{2/L} \cos k x(\tau)~,~c^-_n (\tau)= \epsilon\sqrt{2/L} \sin k x(\tau)$ are effective coupling constants for the accelerated observer coupled to the field \cite{PhysRevD.47.4525, PhysRevD.49.6612}. If we use $\sum_n \to \frac{L}{2 \pi}\int dk$,  $I(k,\tau,\tau')$ becomes \cite{PhysRevD.47.4525, PhysRevD.49.6612},
\begin{equation}
I(k,\tau,\tau')= I(k) \cos k [x(\tau)-x(\tau')]~, \label{IF}
\end{equation}
where $I(k)=\frac{\epsilon^2}{2 \pi \omega}$ is the spectral density of the scalar field seen by an inertial detector. For the case where the scalar field is initially in a vacuum, a scenario that is apt for the present work, the influence of the
quantum field on the detector is expressed in terms of an influence kernel \cite{PhysRevD.47.4525, PhysRevD.49.6612}, appearing in the exponent of the influence functional and has the form,

\begin{eqnarray}
\varsigma (t(\tau), t(\tau'))
&=&\frac{1}{2}\int dk~I(k)~ e^{-i k [x(\tau)-x(\tau')+t(\tau)-t(\tau')]}\nonumber\\
&&+\frac{1}{2}\int dk~I(k)~ e^{-i k [x(\tau')-x(\tau)+t(\tau)-t(\tau')]}\nonumber\\
&=& \frac{1}{2}\int dk~I(k)~ [e^{-2 i k e^{a \Sigma}\sinh(a \Delta)/a}\nonumber\\&&~~~~~~~~~~+e^{-2 i k e^{- a \Sigma}\sinh(a \Delta)/a}]~,
\end{eqnarray}
where $2 \Sigma= \tau+\tau'$, $ \Delta= \tau- \tau'$ and use is made of Eq. (\ref{rindleraccl}). Expanding the exponential terms in the above equations in terms of the Bessel functions of imaginary order, $\varsigma (\tau, \tau')$ can be expressed as,
\begin{eqnarray}\label{eq1}
&&\varsigma (\tau, \tau')\nonumber\\&&=\int^{\infty}_0 dk G(k)\left[\coth\left(\frac{\pi k}{a}\right) \cos k(\tau-\tau')-i \sin k(\tau-\tau')\right]~,\nonumber\\
\end{eqnarray} 
where $G(k)=\frac{2}{\pi a} \sinh(\pi k/a)\int^{\infty}_0 d k' I(k')[K_{2 i k'/a}(2 k' e^{a \Sigma}/a) \\+K_{2 i k'/a}(2 k' e^{-a \Sigma}/a)]$ (for more details, see Appendix \ref{influencekernel}). From Eq. (\ref{eq1}),  the temperature associated with the Unruh effect is seen as $T= \frac{a}{2 \pi k_{B}}$ \cite{PhysRevD.47.4525}. Here $a$ is the acceleration of the detector and $k_{B}$ is the Boltzman constant. This formalism will be applied subsequently.

\section{Hawking radiation for 1+1 dimensional charged black hole}\label{charged black hole}

Here we consider a massless minimal coupled scalar field in a 1+1 dimensional charged black hole (Reissner-Nordstrom black hole). The line element of this black hole is given by \cite{Alvarenga:2003tx, Balbinot:2007kr, Oda:2015mqa, Alvarenga:2003jd, Zhai:2009zz, Liberati:2000sq},
\begin{eqnarray}
ds^2&=&\left(1- \frac{2 M}{r}+\frac{Q^2}{r^2}\right) dt^2- \left(1- \frac{2 M}{r}+\frac{Q^2}{r^2}\right)^{-1} dr^2\nonumber\\
&=&\left(1- \frac{2 M}{r}+\frac{Q^2}{r^2}\right)(dt^2-dr^{{\star}2}),
\end{eqnarray}
where $d r^{\star}=\frac{dr}{\left(1- \frac{2 M}{r}+\frac{Q^2}{r^2}\right)}= \frac{dr}{f(r)}$, $f(r)=\left(1- \frac{2 M}{r}+\frac{Q^2}{r^2}\right)$. Here $M$ is the mass of the black hole and $Q$ is the charge of the black hole. From the equation $f(r)=0$, one can get two horizons as $r_{\pm}=\left(M\pm\sqrt{M^2-Q^2}\right)$, where $r_{\pm}$ are the event horizon and the Cauchy horizons, respectively. Surface gravities at the horizons is defined as $\kappa_{\pm}=\frac{|f'(r)|}{2}|r=r_{\pm}$. Here $\kappa_{+}$ is the surface gravity at the event horizon and $\kappa_{-}$ is the surface gravity at the Cauchy horizon. In terms of the surface gravities and horizons, the tortoise coordinates, $r^{\star}$, can be written as $r^{\star}=\frac{1}{\kappa_{+}}\ln |\frac{r}{r_+}-1|-\frac{1}{\kappa_{-}}\ln |\frac{r}{r_-}-1|$. Now using the expression of surface gravity at the event horizon, we define Kruskal coordinates for the event horizon as,
\begin{equation}
\bar{r}^{\star}(t)=\frac{1}{\kappa_{+}}~e^{\kappa_{+}r^{\star}}\cosh(\kappa_{+}t)~,~\bar{t}(t)=\frac{1}{\kappa_{+}}~e^{\kappa_{+}r^{\star}}\sinh(\kappa_{+}t)~.
\end{equation}
Here, the exponential transformation makes its appearance. Thus, the ordinary coordinate of a charged black hole (t, $r^{\star}$) covers only a wedge-shaped patch of the Kruskal coordinate. A detector at constant $r^{\star}$ is similar to the case of the accelerating observer, as can be seen from an analogy with Eq. (\ref{rindleraccl}). The spectral density, in analogy with Eq. (\ref{IF}), is given by,
\begin{equation}
I(k,t,t')= I(k) \cos k [\bar{r}^{\star}(t)-\bar{r}^{\star}(t')]~,
\end{equation}  
where $I(k)=\frac{\epsilon^2}{2 \pi \omega}$ is the spectral density of the scalar field, as seen by an inertial detector. The influence of the quantum field on the detector is expressed in terms of the influence kernel,
\begin{eqnarray}
&&\varsigma (t,t')
=\frac{1}{2}\int dk~I(k)~ e^{-i k [\bar{r}^{\star}(t)-\bar{r}^{\star}(t')+\bar{t}(t)-\bar{t}(t')]}\nonumber\\
&&\;\;\;\;\;\;\;\;\;\;\;\;+\frac{1}{2}\int dk~I(k)~e^{-i k [\bar{r}^{\star}(t')-\bar{r}^{\star}(t)+\bar{t}(t)-\bar{t}(t')]}\nonumber\\
&=&\int^{\infty}_0 dk~G(k)\left[\coth\left(\frac{\pi k}{\kappa_{+}}\right) \cos k(t-t')-i \sin k(t-t')\right]~,\nonumber\\
\label{eq18}
\end{eqnarray}
where $G(k)=\frac{2}{\pi \kappa_{+}} \sinh(\pi k/\kappa_{+})\int^{\infty}_0 d k' I(k')~\times$\\$[K_{2 i k'/\kappa_{+}}(2 k' e^{\kappa_{+} \Sigma}/\kappa_{+})+K_{2 i k'/\kappa_{+}}(2 k' e^{-\kappa_{+} \Sigma}/\kappa_{+})]$ (for more details, see Appendix \ref{influencekernel}. The calculation is similar but one has to take surface gravity $\kappa_{+}$ instead of acceleration $a$ and appropriate coordinate system).
From Eq. (\ref{eq18}), it can be seen that temperature associated with Hawking radiation for 1+1 dimensional charged black hole is $T=\frac{\kappa_{+}}{2 \pi k_{B}}=\frac{\sqrt{M^2-Q^2}}{2 \pi k_{B} r^2_{+}}$ \cite{PhysRevD.47.4525}. From the expression of the temperature, it follows that for the non-extreme case ($M\neq Q$), temperature is not zero, but it is zero for the extreme case ($M=Q$).

One can also arrive at the result from Eq. (\ref{eq18}), by writing the influence kernel in terms of null coordinates ($U$ and $V$) as \cite{Raval:1995mb},

\begin{eqnarray}\label{hartlestate}
 &&\varsigma (t,t') =\frac{1}{2}\int dk~I(k)~ e^{-i k (U(t)-U(t'))}
 \nonumber\\
&&\;\;\;\;\;\;\;\;\;\;\;\;+\frac{1}{2}\int dk~I(k)~e^{-i k (V(t)-V(t'))}~,
\end{eqnarray}
where $U=\bar{t}-\bar{r}^{\star}=- \frac{1}{\kappa_{+}}~e^{-\kappa_{+}u}$ and $V=\bar{t}+\bar{r}^{\star}=\frac{1}{\kappa_{+}}~e^{\kappa_{+}v} $ are the Kruskal coordinates. Here $u=t-r^{\star}$ and $v=t+r^{\star}$ are the ordinary null coordinates for charged black hole. For a detector at fixed $r^{\star}$ the influence kernel can be written exactly as Eq. (\ref{eq18}). 
One can also consider Unruh vacuum state, using the null coordinates $U$ and $v$, for calculating the Hawking radiation for 1+1 dimensional charged black hole. The influence kernel can be written in terms of null coordinates ($U$ and $v$) as,

\begin{eqnarray}\label{unruhstate}
 &&\varsigma (t,t') =\frac{1}{2}\int dk~I(k)~ e^{-i k (U(t)-U(t'))}
 \nonumber\\
&&\;\;\;\;\;\;\;\;\;\;\;\;+\frac{1}{2}\int dk~I(k)~e^{-i k (v(t)-v(t'))}~.
\end{eqnarray}
 For a detector at fixed $r^{\star}$ the influence kernel (\ref{unruhstate}), using the Unruh vacuum state, can be written as,
\begin{eqnarray}
&&\varsigma (t,t')\nonumber\\&&=\int^{\infty}_0 dk~G(k)\left[\coth\left(\frac{\pi k}{\kappa_{+}}\right) \cos k(t-t')-i \sin k(t-t')\right]\nonumber\\
&&\;\;\;\;\;\;\;\;\;\;\;\;+\frac{1}{2}\int dk~I(k)~e^{-i k (t-t')}~,
\label{unruhtemp}
\end{eqnarray}
where $G(k)=\frac{2}{\pi \kappa_{+}} \sinh(\pi k/\kappa_{+})\int^{\infty}_0 d k' I(k')~\times$\\$[K_{2 i k'/\kappa_{+}}(2 k' e^{-\kappa_{+} \Sigma}/\kappa_{+})]$ (for more details, see Appendix \ref{unruhsate}). As $v$ is not same as $V$, we get a form different from that of Eq. (\ref{eq18}).
From Eq. (\ref{unruhtemp}), it can be seen that temperature associated with Hawking radiation for 1+1 dimensional charged black hole is $T=\frac{\kappa_{+}}{2 \pi k_{B}}=\frac{\sqrt{M^2-Q^2}}{2 \pi k_{B} r^2_{+}}$.  Thus, we get the same temperature using the Unruh vacuum state instead of the Hartle-Hawking vacuum state.

\section{Hawking radiation from the Kerr black hole} \label{kerr black hole}

Here we consider a massless minimal coupled scalar field in Kerr black holes. Near the horizon, the scalar field theory in a 4-$d$ Kerr black hole spacetime can be reduced to the 2-$d$ field theory \cite{PhysRevD.74.044018}. 

In Boyer-Lindquist coordinates, Kerr metric is given by \cite{doi:10.1063/1.1705193, Kerr:2007dk, Heinicke:2014ipp, KRASINSKI197822, Teukolsky:2014vca, Visser:2007fj, Smailagic:2010nv, Dadhich:2013qx},
\begin{multline}
 ds^2 = -\frac{\Delta-a^2 \sin^2 \theta}{\Sigma}dt^2 - 2a\sin^2 \theta
  \frac{r^2+a^2-\Delta}{\Sigma}dtd\phi\\
  + \frac{(r^2+a^2)^2-\Delta a^2
  \sin^2 \theta}{\Sigma}\sin^2 \theta d\phi^2 +
  \frac{\Sigma}{\Delta}dr^2 + \Sigma d\theta^2~, 
\end{multline}
where 
\begin{equation}
 \begin{split}
  \Sigma &= r^2 + a^2 \cos^2 \theta \ , \\
  \Delta &= r^2 - 2Mr + a^2 \\
         &= (r-r_+)(r-r_-)\ .
 \end{split}
\end{equation}
Here $M$ is the mass of the black hole, $a$ is the angular momentum per unit mass of the black hole, and $r=r_{\pm}$ is the event horizon and the Cauchy horizon, 
respectively.
The determinant of the above metric is
\begin{equation}
 \sqrt{-g} = \Sigma \sin \theta  \ ,
\end{equation}
and the inverse of the above metric of ($t,\phi$) parts is given by \cite{PhysRevD.74.044018},
\begin{equation}
 \begin{split}
  g^{tt} &= -\frac{(r^2+a^2)^2-\Delta a^2 \sin^2\theta}{\Sigma\Delta}  \ ,\\
  g^{\phi\phi} &=
  \frac{\Delta-a^2\sin^2\theta}{\Sigma\Delta\sin^2\theta}  \ , \\
  g^{t\phi} &= -\frac{a(r^2+a^2-\Delta)}{\Sigma\Delta}  \ .
 \end{split}
\end{equation}
The action for a massless minimal coupled scalar field in the 4-$d$ Kerr spacetime is given by \cite{PhysRevD.74.044018},
\begin{equation}
\begin{split}
 S[\varphi] =& \frac{1}{2}\int d^4 x \sqrt{-g}\,\varphi \nabla^2 \varphi \\
=& \frac{1}{2}\int d^4 x\sqrt{-g} \,\varphi\frac{1}{\Sigma}
 \left[-\left(\frac{(r^2+a^2)^2}{\Delta}-a^2 \sin^2 \theta
 \right)\partial_t^2 \right.\\
 &- \frac{2a(r^2 + a^2 - \Delta)}{\Delta}\partial_t
 \partial_\phi + \left(\frac{1}{\sin^2
 \theta}-\frac{a^2}{\Delta}\right)\partial_\phi^2\\
 &+ \left. \partial_r \Delta
 \partial_r + \frac{1}{\sin\theta}\partial_\theta \sin \theta
 \partial_\theta \right]\varphi  \ .
\end{split}
\end{equation}
After taking the limit $r \rightarrow r_+$ and the leading order terms, we get,
\begin{equation}
\begin{split}
 S[\varphi] =& \frac{1}{2}\int d^4x\sin\theta \,\varphi
 \left[ -\frac{(r_+^2+a^2)^2}{\Delta}\partial_t^2 \right.\\
 &- \left.\frac{2a(r_+^2 +
 a^2)}{\Delta}\partial_t \partial_\phi
 -\frac{a^2}{\Delta}\partial_\phi^2
 + \partial_r \Delta \partial_r
 \right]\varphi  \ .
\label{eq:dalembertian}
\end{split}
\end{equation}
Now we transform the coordinates to the locally non-rotating coordinate system, given by \cite{PhysRevD.74.044018},
\begin{equation}
\begin{cases}
 \psi = \phi - \Omega_H t \ , \\
 \xi = t  \ ,
\end{cases}
\end{equation}
where
\begin{equation}
 \Omega_H \equiv \frac{a}{r_+^2 + a^2}  \ .
\end{equation}
 Using $(\xi,r,\theta,\psi)$ coordinates, the action (\ref{eq:dalembertian}) can be rewritten as \cite{PhysRevD.74.044018},
\begin{equation}
 S[\varphi] = \frac{a}{2\Omega_H}\int d^4x \sin\theta \,\varphi \left( -\frac{1}{f(r)}\partial_\xi^2 + \partial_r f(r) \partial_r \right)\varphi \ ,
\end{equation}
where 
\begin{equation}
 f(r) \equiv \frac{\Omega_H \Delta}{a} \ .
\end{equation}
Applying 
the spherical harmonics expansion 
$\varphi(x) = \sum_{l,m} \varphi_{l\,m}(\xi,r)Y_{l\,m}(\theta,\psi)$, 
 finally one can get the effective 2-dimensional action as \cite{PhysRevD.74.044018},
\begin{equation}
\begin{split}
 S[\varphi] =& \frac{a}{\Omega_H}\sum_{l,m}\frac{1}{2}\int d\xi dr\\
&\times \varphi_{l\,m}
  \left(-\frac{1}{f(r)}\partial_\xi^2 + \partial_r f(r) \partial_r
  \right) \varphi_{l\,m}  \ .
\end{split}
\end{equation}
The effective 2-dimensional metric from the above action can be expressed
as \cite{PhysRevD.74.044018}, 
\begin{equation}
 ds^2 = -f(r)d\xi^2 + \frac{1}{f(r)}dr^2 \ .
\end{equation}
Hence, in the near-horizon region, the
geometry of Kerr spacetime is the same as the Rindler spacetime when $r_+ > r_-$.
One can rewrite this metric as,
\begin{eqnarray}
ds^2 =f(r)(- d\xi^2+dr^{{\star}2}),
\end{eqnarray}
where $d r^{\star}= \frac{dr}{f(r)}$. Surface gravity at the event horizon is defined as $\kappa_{+}=\frac{|f'(r)|}{2}|r=r_{+}$. Now using the expression of the surface gravity, we define Kruskal coordinates as,

\begin{equation}
\bar{r}^{\star}(\xi)=\frac{1}{\kappa_{+}}~e^{\kappa_{+} r^{\star}}\cosh(\kappa_{+} \xi)~,~\bar{t}(\xi)=\frac{1}{\kappa_{+} }~e^{\kappa_{+} r^{\star}}\sinh(\kappa_{+} \xi)~.
\end{equation}
Here we again come across the exponential transformation relation. Thus, in the near-horizon region, the ordinary coordinate of Kerr black hole ($\xi$, $r^{\star}$) covers only a wedge-shaped patch of the Kruskal coordinate.
For a detector at constant $r^{\star}, $ the case is similar to an accelerating observer, as can be seen from an analogy with Eq. (\ref{rindleraccl}). The spectral density, in analogy with Eq. (\ref{IF}), is given by,

\begin{equation}
I(k,\xi,\xi')= I(k) \cos k [\bar{r}^{\star}(\xi)-\bar{r}^{\star}(\xi')]~,
\end{equation}  
where $I(k)=\frac{\epsilon^2}{2 \pi \omega}$ is the spectral density of the scalar field seen by an inertial detector. The influence of the quantum field on the detector is expressed in terms of the influence
kernel, having the form 
\begin{eqnarray}
&&\varsigma (\xi,\xi')
=\frac{1}{2}\int dk~I(k)~ e^{-i k [\bar{r}^{\star}(\xi)-\bar{r}^{\star}(\xi')+\bar{t}(\xi)-\bar{t}(\xi')]}\nonumber\\
&&\;\;\;\;\;\;\;\;\;\;\;\;+\frac{1}{2}\int dk~I(k)~e^{-i k [\bar{r}^{\star}(\xi')-\bar{r}^{\star}(\xi)+\bar{t}(\xi)-\bar{t}(\xi')]}\nonumber\\
&=&\int^{\infty}_0 dk~G(k)\left[\coth\left(\frac{\pi k}{\kappa_{+}}\right) \cos k(\xi-\xi')-i \sin k(\xi-\xi')\right]~,\nonumber\\
\label{eq34}
\end{eqnarray}
where $G(k)=\frac{2}{\pi \kappa_{+}} \sinh(\pi k/\kappa_{+})\int^{\infty}_0 d k' I(k')~\times$\\$[K_{2 i k'/\kappa_{+}}(2 k' e^{\kappa_{+} \Sigma}/\kappa_{+})+K_{2 i k'/\kappa_{+}}(2 k' e^{-\kappa_{+} \Sigma}/\kappa_{+})]$ (for more details, see Appendix \ref{influencekernel}, where one has to take surface gravity $\kappa_{+}$ instead of acceleration $a$ and appropriate coordinate system).
From Eq. (\ref{eq34}), it can be seen that the temperature associated with Hawking radiation for Kerr black hole  is $T=\frac{\kappa_{+}}{2 \pi k_{B}}=\frac{1}{2 \pi k_{B}} \left(\frac{\sqrt{M^2-a^2}}{M+\sqrt{M^2-a^2}}\right)$. From this it follows that for the non-extreme case ($M\neq a$), the temperature is not zero, but it is zero for the extreme case ($M=a$). 

 As shown above, one can write the influence kernel interms of null coordinates ($U$ and $V$) as \cite{Raval:1995mb},

\begin{eqnarray}\label{hartlestate1}
 &&\varsigma (\xi,\xi') =\frac{1}{2}\int dk~I(k)~ e^{-i k (U(\xi)-U(\xi'))}
 \nonumber\\
&&\;\;\;\;\;\;\;\;\;\;\;\;+\frac{1}{2}\int dk~I(k)~e^{-i k (V(\xi)-V(\xi'))}~,
\end{eqnarray}
where $U=\bar{t}-\bar{r}^{\star}=- \frac{1}{\kappa_{+}}~e^{-\kappa_{+}u}$ and $V=\bar{t}+\bar{r}^{\star}=\frac{1}{\kappa_{+}}~e^{\kappa_{+}v} $ are the Kruskal coordinates and $u=\xi-r^{\star}$ and $v=\xi+r^{\star}$ are ordinary null coordinates for Kerr black hole in the near-horizon region. For a detector at fixed $r^{\star}$ the influence kernel can be written exactly as Eq. (\ref{eq34}). 
One can also consider the Unruh vacuum state for calculating the Hawking radiation for the Kerr black hole. The influence kernel can be written as in terms of null coordinates ($U$ and $v$) as,

\begin{eqnarray}\label{unruhstate1}
 &&\varsigma (\xi,\xi') =\frac{1}{2}\int dk~I(k)~ e^{-i k (U(\xi)-U(\xi'))}
 \nonumber\\
&&\;\;\;\;\;\;\;\;\;\;\;\;+\frac{1}{2}\int dk~I(k)~e^{-i k (v(\xi)-v(\xi'))}~.
\end{eqnarray}
 For a detector at fixed $r^{\star}$ the influence kernel (\ref{unruhstate1}), using the Unruh vacuum state, can be written as,
\begin{eqnarray}
&&\varsigma (\xi,\xi')\nonumber\\&&=\int^{\infty}_0 dk~G(k)\left[\coth\left(\frac{\pi k}{\kappa_{+}}\right) \cos k(\xi-\xi')-i \sin k(\xi-\xi')\right]\nonumber\\
&&\;\;\;\;\;\;\;\;\;\;\;\;+\frac{1}{2}\int dk~I(k)~e^{-i k (\xi-\xi')}~,
\label{unruhtemp1}
\end{eqnarray}
where $G(k)=\frac{2}{\pi \kappa_{+}} \sinh(\pi k/\kappa_{+})\int^{\infty}_0 d k' I(k')~\times$\\$[K_{2 i k'/\kappa_{+}}(2 k' e^{-\kappa_{+} \Sigma}/\kappa_{+})]$ (for more details, see Appendix \ref{unruhsate}, where one has to consider appropriate coordinate system).
From Eq. (\ref{unruhtemp1}), it can be seen that the temperature associated with Hawking radiation for Kerr black hole is $T=\frac{\kappa_{+}}{2 \pi k_{B}}=\frac{1}{2 \pi k_{B}} \left(\frac{\sqrt{M^2-a^2}}{M+\sqrt{M^2-a^2}}\right)$. We get the same temperature using the Unruh vacuum state instead of the Hartle-Hawking vacuum state.

\section{Hawking radiation for 1+1
DIMENSIONAL Schwarzschild de-Sitter spacetime} \label{Schwarschild de-Sitter spacetime}

We now consider a massless minimal coupled scalar field in the Schwarzschild-de Sitter spacetime. The line element for this spacetime in $1+1 d$ is \cite{Bhattacharya:2013tq, Shankaranarayanan:2003ya, PhysRevD.66.124009, Choudhury:2004ph, Pappas:2017kam, Robson:2019yzx, RAHMAN20121},
\begin{equation}\label{eq2}
ds^2=\left(1-\frac{2 M}{r}-\frac{\Lambda r^2}{3}\right)dt^2-\left(1-\frac{2 M}{r}-\frac{\Lambda r^2}{3}\right)^{-1} dr^2.
\end{equation}
Here $M$ is the mass of the black hole, and $\Lambda$ is the positive cosmological constant.
For $3 M \sqrt{\Lambda}<1$, this spacetime has three Killing horizons, which are,

\begin{eqnarray}
r_{H}&=&\frac{2}{\sqrt{\Lambda}}~\cos\left[\frac{1}{3}\cos^{-1}(3 M \sqrt{\Lambda})+\frac{\pi}{3}\right],\nonumber\\
r_{c}&=&\frac{2}{\sqrt{\Lambda}}~\cos\left[\frac{1}{3}\cos^{-1}(3 M \sqrt{\Lambda})-\frac{\pi}{3}\right],\nonumber\\
r_{u}&=&-(r_{H}+r_{c})~.
\end{eqnarray}
$r_{H}$ is the black hole event horizon and $r_{c}>r_{H}$ is the cosmological horizon, and $r_{u}$ is negative, which is the unphysical horizon. Now in terms of tortoise coordinates, the above line element (\ref{eq2}) can be written as,

\begin{equation}
ds^2=\left(1-\frac{2 M}{r}-\frac{\Lambda r^2}{3}\right)(dt^2-dr^{{\star}^2}).
\end{equation}
The exact form of tortoise coordinate, $r^{\star}$, is given below,
\begin{eqnarray}
r^{\star}&=&\int \frac{dr}{(1-\frac{2 M}{r}-\frac{\Lambda r^2}{3})}\nonumber\\
&=&\frac{1}{\kappa_{H}}\ln\left|\frac{r}{r_{H}}-1\right|-\frac{1}{\kappa_{c}}\ln\left|\frac{r}{r_{c}}-1\right|+\frac{1}{\kappa_{u}}\ln\left|\frac{r}{r_{u}}-1\right|~,\nonumber
\end{eqnarray}
where $\kappa_{i}$ is the surface gravity of the corresponding horizon $r_{i}$ which is given by, 
\begin{equation}
\kappa_{i}=\frac{1}{2}\left|\partial_r\left(1-\frac{2 M}{r}-\frac{\Lambda r^2}{3}\right)\right|_{r=r_{i}}~.
\end{equation}
Here $\kappa_{H}$ is the surface gravity at the black hole event horizon, and $\kappa_{c}$ is the surface gravity at the cosmological horizon.
We now define Kruskal coordinates for the black hole event horizon as \cite{Bhattacharya:2013tq},
\begin{equation}
\bar{r}^{\star}(t)=\frac{1}{\kappa_{H}}~e^{\kappa_{H}r^{\star}}\cosh(\kappa_{H}t)~,~\bar{t}(t)=\frac{1}{\kappa_{H}}~e^{\kappa_{H}r^{\star}}\sinh(\kappa_{H}t)~.
\end{equation}
We also define Kruskal coordinates for the cosmological horizon as \cite{Bhattacharya:2013tq},
\begin{equation}
\bar{r}^{{\star}'}(t)=\frac{1}{\kappa_{c}}~e^{\kappa_{c}r^{\star}}\cosh(\kappa_{c}t)~,~\bar{t'}(t)=\frac{1}{\kappa_{c}}~e^{\kappa_{c}r^{\star}}\sinh(\kappa_{c}t)~.
\end{equation}
For both the cases we have exponential transformation relations. Thus, the ordinary coordinate of Schwarzschild-de Sitter spacetime (t, $r^{\star}$) covers only a wedge-shaped patch of the Kruskal coordinate. It follows that
for a detector at constant $r^{\star}$, the case is similar to the accelerating observer. The spectral density for black hole event horizon, in analogy to the previous cases, is, 

\begin{equation}
I(k,\tau,\tau')= I(k) \cos k [\bar{r}^{\star}(t)-\bar{r}^{\star}(t')]~,
\end{equation}  
where $I(k)=\frac{\epsilon^2}{2 \pi \omega}$ is the spectral density of the scalar field seen by an inertial detector. The influence kernel characterizing the influence of the quantum field on the detector has the form
\begin{eqnarray}
&&\varsigma (t,t')
=\frac{1}{2}\int dk~I(k)~ e^{-i k [\bar{r}^{\star}(t)-\bar{r}^{\star}(t')+\bar{t}(t)-\bar{t}(t')]}\nonumber\\
&&\;\;\;\;\;\;\;\;\;\;\;\;+\frac{1}{2}\int dk~I(k)~e^{-i k [\bar{r}^{\star}(t')-\bar{r}^{\star}(t)+\bar{t}(t)-\bar{t}(t')]}\nonumber\\
&=&\int^{\infty}_0 dk~G(k)\left[\coth\left(\frac{\pi k}{\kappa_{H}}\right) \cos k(t-t')-i \sin k(t-t')\right]~,\nonumber\\
\label{eq43}
\end{eqnarray}
where $G(k)=\frac{2}{\pi \kappa_{H}} \sinh(\pi k/\kappa_{H})\int^{\infty}_0 d k' I(k')~\times$\\$[K_{2 i k'/\kappa_{H}}(2 k' e^{\kappa_{H} \Sigma}/\kappa_{H})+K_{2 i k'/\kappa_{H}}(2 k' e^{-\kappa_{H} \Sigma}/\kappa_{H})]$ (for more details, see Appendix \ref{influencekernel}. The calculation is similar but one has to take surface gravity $\kappa_{H}$ instead of acceleration $a$ and appropriate coordinate system).
From Eq. (\ref{eq43}), the temperature associated with Hawking radiation for black hole horizon is seen to be $T_{H}=\frac{\kappa_{H}}{2 \pi k_{B}}$.

Similarly, the spectral density for cosmological horizon is given by, 

\begin{equation}
I(k,\tau,\tau')= I(k) \cos k [\bar{r}^{{\star}'}(t)-\bar{r}^{{\star}'}(t')]~,
\end{equation}  
where $I(k)=\frac{\epsilon^2}{2 \pi \omega}$ is the spectral density of the scalar field. The form of the influence kernel is 
\begin{eqnarray}
&&\varsigma (t,t')
=\frac{1}{2}\int dk~I(k)~ e^{-i k [\bar{r}^{{\star}'}(t)-\bar{r}^{{\star}'}(t')+\bar{t'}(t)-\bar{t'}(t')]}\nonumber\\
&&\;\;\;\;\;\;\;\;\;\;\;\;+\frac{1}{2}\int dk~I(k)~e^{-i k [\bar{r}^{{\star}'}(t')-\bar{r}^{{\star}'}(t)+\bar{t'}(t)-\bar{t'}(t')]}\nonumber\\
&=&\int^{\infty}_0 dk~G(k)\left[\coth\left(\frac{\pi k}{\kappa_{c}}\right) \cos k(t-t')-i \sin k(t-t')\right]~,\nonumber\\
\label{eq45}
\end{eqnarray}
where $G(k)=\frac{2}{\pi \kappa_{c}} \sinh(\pi k/\kappa_{c})\int^{\infty}_0 d k' I(k')~\times$\\$[K_{2 i k'/\kappa_{c}}(2 k' e^{\kappa_{c} \Sigma}/\kappa_{c})+K_{2 i k'/\kappa_{c}}(2 k' e^{-\kappa_{c} \Sigma}/\kappa_{c})]$ (for more details, see Appendix \ref{influencekernel}, with surface gravity $\kappa_{c}$ instead of acceleration $a$ and appropriate coordinate system).
From Eq. (\ref{eq45}), the temperature associated with Hawking radiation for the cosmological horizon comes out to be $T_{c}=\frac{\kappa_{c}}{2 \pi k_{B}}$. 

The influence kernel can be expressed in terms of null coordinates ($U$ and $V$) as \cite{Raval:1995mb},

\begin{eqnarray}\label{hartlestate2}
 &&\varsigma (t,t') =\frac{1}{2}\int dk~I(k)~ e^{-i k (U(t)-U(t'))}
 \nonumber\\
&&\;\;\;\;\;\;\;\;\;\;\;\;+\frac{1}{2}\int dk~I(k)~e^{-i k (V(t)-V(t'))}~,
\end{eqnarray}
where $U=\bar{t}-\bar{r}^{\star}=- \frac{1}{\kappa_{H}}~e^{-\kappa_{H}u}$ and $V=\bar{t}+\bar{r}^{\star}=\frac{1}{\kappa_{H}}~e^{\kappa_{H}v} $ are the Kruskal coordinates for the black
hole event horizon and $u=t-r^{\star}$ and $v=t+r^{\star}$ are ordinary null coordinates for  Schwarschild de-Sitter spacetime. For the Unruh vacuum state the influence kernel, using the null coordinates $U$ and $v$, is

\begin{eqnarray}\label{unruhstate2}
 &&\varsigma (t,t') =\frac{1}{2}\int dk~I(k)~ e^{-i k (U(t)-U(t'))}
 \nonumber\\
&&\;\;\;\;\;\;\;\;\;\;\;\;+\frac{1}{2}\int dk~I(k)~e^{-i k (v(t)-v(t'))}~.
\end{eqnarray}
 At fixed $r^{\star}$ Eq. (\ref{unruhstate2}) becomes,
\begin{eqnarray}
&&\varsigma (t,t')\nonumber\\&&=\int^{\infty}_0 dk~G(k)\left[\coth\left(\frac{\pi k}{\kappa_{H}}\right) \cos k(t-t')-i \sin k(t-t')\right]\nonumber\\
&&\;\;\;\;\;\;\;\;\;\;\;\;+\frac{1}{2}\int dk~I(k)~e^{-i k (t-t')}~,
\label{unruhtemp2}
\end{eqnarray}
where $G(k)=\frac{2}{\pi \kappa_{H}} \sinh(\pi k/\kappa_{H})\int^{\infty}_0 d k' I(k')~\times$\\$[K_{2 i k'/\kappa_{H}}(2 k' e^{-\kappa_{H} \Sigma}/\kappa_{H})]$ (for more details, see Appendix \ref{unruhsate}, where one has to consider appropriate coordinate system and surface gravity $\kappa_{H}$ instead of $\kappa_{+}$).
From Eq. (\ref{unruhtemp2}), it can be seen that temperature associated with Hawking radiation for  black hole horizon is $T_{H}=\frac{\kappa_{H}}{2 \pi k_{B}}$, same as that using the Hartle-Hawking vacuum state.

Similarly, one can also consider Unruh vacuum state for calculating the Hawking radiation for  cosmological horizon. In this case, the temperature associated with Hawking radiation for cosmological horizon is $T_{H}=\frac{\kappa_{c}}{2 \pi k_{B}}$ which is the same as that using the Hartle-Hawking vacuum state.

\section{Hawking radiation for 1+1
DIMENSIONAL Schwarzschild anti-de Sitter spacetime} \label{Schwarschild anti-de Sitter spacetime}

Now we consider a massless minimal coupled scalar field in $1+1~ d$ Schwarzschild anti-de Sitter spacetime. The line element for this spacetime is \cite{Socolovsky:2017nff, Hawking:1982dh, Robson:2019wfu, Lanteri:2020trb, book:15291, AtiqurRahman:2012km}

\begin{eqnarray}
ds^2&=&\left(1- \frac{2 M}{r}+\frac{r^2}{l^2}\right) dt^2- \left(1- \frac{2 M}{r}+\frac{r^2}{l^2}\right)^{-1} dr^2\nonumber\\
&=&\left(1- \frac{2 M}{r}+\frac{r^2}{l^2}\right)(dt^2-dr^{{\star}2}),
\end{eqnarray}
where $d r^{\star}=\frac{dr}{\left(1- \frac{2 M}{r}+\frac{r^2}{l^2}\right)}= \frac{dr}{f(r)}$, $f(r)=\left(1- \frac{2 M}{r}+\frac{r^2}{l^2}\right)$. Here $M$ is the mass of the black hole and $l^2$ is connected with the positive cosmological constant. From the equation $f(r)=0$, 
one can get one horizon as, $r_{+}=\frac{2}{3}\sqrt{3}l \sinh\left(\frac{1}{3}\sqrt{3}l \sinh^{-1}\left(3 \sqrt{3}\frac{m}{l}\right)\right)$, where $r_{+}$ is the event horizon. Surface gravity at the event horizon is defined as $\kappa=\frac{|f'(r)|}{2}|r=r_{+}$. We define the Kruskal coordinates as,

\begin{equation}
\bar{r}^{\star}(t)=\frac{1}{\kappa}~e^{\kappa r^{\star}}\cosh(\kappa t)~,~\bar{t}(t)=\frac{1}{\kappa }~e^{\kappa r^{\star}}\sinh(\kappa t)~,
\end{equation}
an exponential transformation. It follows that the ordinary coordinate of Schwarzschild anti-de Sitter (t, $r^{\star}$) covers only a wedge-shaped patch of the Kruskal coordinate.
For a detector at constant $r^{\star},$ the case is  similar to an accelerating observer. The spectral density is given by,

\begin{equation}
I(k,\tau,\tau')= I(k) \cos k [\bar{r}^{\star}(t)-\bar{r}^{\star}(t')]~,
\end{equation}  
where $I(k)=\frac{\epsilon^2}{2 \pi \omega}$. The influence of the quantum field on the detector is expressed in terms of an influence
kernel 
\begin{eqnarray}
&&\varsigma (t,t')=\frac{1}{2}\int dk~I(k)~ e^{-i k [\bar{r}^{\star}(t)-\bar{r}^{\star}(t')+\bar{t}(t)-\bar{t}(t')]}\nonumber\\
&&\;\;\;\;\;\;\;\;\;\;\;\;+\frac{1}{2}\int dk~I(k)~e^{-i k [\bar{r}^{\star}(t')-\bar{r}^{\star}(t)+\bar{t}(t)-\bar{t}(t')]}\nonumber\\
&=&\int^{\infty}_0 dk~G(k)\left[\coth\left(\frac{\pi k}{\kappa}\right) \cos k(t-t')-i \sin k(t-t')\right]~,\nonumber\\
\label{eq49}
\end{eqnarray}
where $G(k)=\frac{2}{\pi \kappa} \sinh(\pi k/\kappa)\int^{\infty}_0 d k' I(k')[K_{2 i k'/\kappa} (2 k' e^{\kappa \Sigma}/\kappa)\\+K_{2 i k'/\kappa}(2 k' e^{-\kappa \Sigma}/\kappa)]$ (for more details, see Appendix \ref{influencekernel}, with surface gravity $\kappa$ instead of acceleration $a$ and appropriate coordinate system).
From Eq. (\ref{eq49}), it is seen that the temperature associated with Hawking radiation for $1+1~ d$ Schwarzschild anti-de Sitter spacetime  is $T=\frac{\kappa }{2 \pi k_{B}}=\frac{1}{2 \pi k_{B}} \left(\frac{m}{r^2_+}+\frac{r_+}{l^2}\right)$.

The influence kernel in terms of null coordinates ($U$ and $V$) is \cite{Raval:1995mb},

\begin{eqnarray}\label{hartlestate4}
 &&\varsigma (t,t') =\frac{1}{2}\int dk~I(k)~ e^{-i k (U(t)-U(t'))}
 \nonumber\\
&&\;\;\;\;\;\;\;\;\;\;\;\;+\frac{1}{2}\int dk~I(k)~e^{-i k (V(t)-V(t'))}~,
\end{eqnarray}
where $U=\bar{t}-\bar{r}^{\star}=- \frac{1}{\kappa}~e^{-\kappa u}$ and $V=\bar{t}+\bar{r}^{\star}=\frac{1}{\kappa}~e^{\kappa v} $ are the Kruskal coordinates and $u=t-r^{\star}$ and $v=t+r^{\star}$ are ordinary null coordinates for  Schwarschild anti-de sitter spacetime. For a detector at fixed $r^{\star}$ the influence kernel can be written exactly as Eq. (\ref{eq49}). 
One can also consider Unruh vacuum state for calculating the Hawking radiation for $1+1~ d$ Schwarzschild anti-de Sitter spacetime. The influence kernel in terms of null coordinates ($U$ and $v$) is,

\begin{eqnarray}\label{unruhstate4}
 &&\varsigma (t,t') =\frac{1}{2}\int dk~I(k)~ e^{-i k (U(t)-U(t'))}
 \nonumber\\
&&\;\;\;\;\;\;\;\;\;\;\;\;+\frac{1}{2}\int dk~I(k)~e^{-i k (v(t)-v(t'))}~.
\end{eqnarray}
 For a detector at fixed $r^{\star}$ the influence kernel (\ref{unruhstate4}) using Unruh vacuum state can be written as,
\begin{eqnarray}
&&\varsigma (t,t')\nonumber\\&&=\int^{\infty}_0 dk~G(k)\left[\coth\left(\frac{\pi k}{\kappa}\right) \cos k(t-t')-i \sin k(t-t')\right]\nonumber\\
&&\;\;\;\;\;\;\;\;\;\;\;\;+\frac{1}{2}\int dk~I(k)~e^{-i k (t-t')}~,
\label{unruhtemp4}
\end{eqnarray}
where $G(k)=\frac{2}{\pi \kappa} \sinh(\pi k/\kappa)\int^{\infty}_0 d k' I(k')~\times$\\$[K_{2 i k'/\kappa}(2 k' e^{-\kappa \Sigma}/\kappa)]$ (for more details, see Appendix \ref{unruhsate}, with an appropriate coordinate system and surface gravity $\kappa$ instead of $\kappa_{+}$).
From Eq. (\ref{unruhtemp4}), it can be seen that temperature associated with Hawking radiation for $1+1 d$ Schwarzschild anti-de Sitter spacetime is $T_{H}=\frac{\kappa}{2 \pi k_{B}}=\frac{1}{2 \pi k_{B}} \left(\frac{m}{r^2_+}+\frac{r_+}{l^2}\right)$, as obtained  using the Hartle-Hawking vacuum state.

\section{Hawking radiation for 1+1
DIMENSIONAL Reissner-Nordstrom de-Sitter spacetime} \label{Reissner-Nordstrom de-Sitter}

We now consider a massless minimal coupled scalar field in the Reissner-Nordstrom de-Sitter spacetime. The line element for this spacetime in $1+1~d$ is \cite{Li:2021axp, Zhang:2016nws, Hollands:2019whz,Guo:2005hw, Ahmed:2016lou},
\begin{eqnarray}\label{eq02}
ds^2&=&\left(1-\frac{2 M}{r}+\frac{Q^2}{r^2}-\frac{\Lambda r^2}{3}\right)dt^2\nonumber\\&-&\left(1-\frac{2 M}{r} +\frac{Q^2}{r^2} -\frac{\Lambda r^2}{3}\right)^{-1} dr^2\nonumber\\
&=&\left(1- \frac{2 M}{r}+\frac{Q^2}{r^2}-\frac{\Lambda r^2}{3}\right)(dt^2-dr^{{\star}2}),
\end{eqnarray}
where $d r^{\star}=\frac{dr}{\left(1- \frac{2 M}{r}+\frac{Q^2}{r^2}-\frac{\Lambda r^2}{3}\right)}= \frac{dr}{f(r)}$, $f(r)=\left(1- \frac{2 M}{r}+\frac{Q^2}{r^2}-\frac{\Lambda r^2}{3}\right)$.
Here $M$ is the mass of the black hole, $Q$ is the charge of the black hole, and $\Lambda$ is the positive cosmological constant.
From the equation $f(r)=0$, one can get three horizons for this spaetime. Here we consider $r_{\pm}$ as the event and the Cauchy horizons, respectively, and $r_{c}$ the cosmological horizon. Surface gravities at the horizons are defined as $\kappa_{\pm}=\frac{|f'(r)|}{2}|_{r=r_{\pm}}$ and $\kappa_{c}=\frac{|f'(r)|}{2}|_{r=r_{c}}$. Here $\kappa_{+}$ is the surface gravity at the event horizon, $\kappa_{-}$ is the surface gravity at the Cauchy horizon, and $\kappa_{c}$ is the surface gravity at the cosmological horizon, respectively. The exact form of tortoise coordinate, $r^{\star}$, in terms of the surface gravities and the horizons is,
\begin{eqnarray}
r^{\star}&=&\int \frac{dr}{(1-\frac{2 M}{r}+\frac{Q^2}{r^2}-\frac{\Lambda r^2}{3})}\nonumber\\
&=&\frac{1}{\kappa_{+}}\ln\left|\frac{r}{r_{+}}-1\right|- \frac{1}{\kappa_{-}}\ln\left|\frac{r}{r_{-}}-1\right|\nonumber\\
&-&\frac{1}{\kappa_{c}}\ln\left|\frac{r}{r_{c}}-1\right|+\frac{1}{\kappa_{u}}\ln\left|\frac{r}{r_{u}}-1\right|~,\nonumber
\end{eqnarray}
where $r_{u}$ is negative, which is the unphysical horizon.
We now define Kruskal coordinates for the event horizon as,
\begin{equation}
\bar{r}^{\star}(t)=\frac{1}{\kappa_{+}}~e^{\kappa_{+}r^{\star}}\cosh(\kappa_{+}t)~,~\bar{t}(t)=\frac{1}{\kappa_{+}}~e^{\kappa_{+}r^{\star}}\sinh(\kappa_{+}t)~.
\end{equation}
We also define Kruskal coordinates for the cosmological horizon as,
\begin{equation}
\bar{r}^{{\star}'}(t)=\frac{1}{\kappa_{c}}~e^{\kappa_{c}r^{\star}}\cosh(\kappa_{c}t)~,~\bar{t'}(t)=\frac{1}{\kappa_{c}}~e^{\kappa_{c}r^{\star}}\sinh(\kappa_{c}t)~.
\end{equation}
For both the cases we have exponential transformation relations. Thus, the ordinary coordinate of Reissner-Nordstrom de-Sitter spacetime (t, $r^{\star}$) covers only a wedge-shaped patch of the Kruskal coordinate. It follows that
for a detector at constant $r^{\star}$, the case is similar to the accelerating observer. The spectral density for the event horizon, in analogy to the previous cases, is, 

\begin{equation}
I(k,\tau,\tau')= I(k) \cos k [\bar{r}^{\star}(t)-\bar{r}^{\star}(t')]~,
\end{equation}  
where $I(k)=\frac{\epsilon^2}{2 \pi \omega}$ is the spectral density of the scalar field seen by an inertial detector. The influence kernel characterizing the influence of the quantum field on the detector has the form
\begin{eqnarray}
&&\varsigma (t,t')
=\frac{1}{2}\int dk~I(k)~ e^{-i k [\bar{r}^{\star}(t)-\bar{r}^{\star}(t')+\bar{t}(t)-\bar{t}(t')]}\nonumber\\
&&\;\;\;\;\;\;\;\;\;\;\;\;+\frac{1}{2}\int dk~I(k)~e^{-i k [\bar{r}^{\star}(t')-\bar{r}^{\star}(t)+\bar{t}(t)-\bar{t}(t')]}\nonumber\\
&=&\int^{\infty}_0 dk~G(k)\left[\coth\left(\frac{\pi k}{\kappa_{+}}\right) \cos k(t-t')-i \sin k(t-t')\right]~,\nonumber\\
\label{eq430}
\end{eqnarray}
where $G(k)=\frac{2}{\pi \kappa_{+}} \sinh(\pi k/\kappa_{+})\int^{\infty}_0 d k' I(k')~\times$\\$[K_{2 i k'/\kappa_{+}}(2 k' e^{\kappa_{+} \Sigma}/\kappa_{+})+K_{2 i k'/\kappa_{+}}(2 k' e^{-\kappa_{+} \Sigma}/\kappa_{+})]$ (for more details, see Appendix \ref{influencekernel}, with surface gravity $\kappa_{+}$ replacing acceleration $a$ and an appropriate coordinate system).
From Eq. (\ref{eq430}), the temperature associated with Hawking radiation for black hole horizon is seen to be $T_{+}=\frac{\kappa_{+}}{2 \pi k_{B}}$.

Similarly, the spectral density for cosmological horizon is given by, 

\begin{equation}
I(k,\tau,\tau')= I(k) \cos k [\bar{r}^{{\star}'}(t)-\bar{r}^{{\star}'}(t')]~,
\end{equation}  
where $I(k)=\frac{\epsilon^2}{2 \pi \omega}$ is the spectral density of the scalar field. The form of the influence kernel is 
\begin{eqnarray}
&&\varsigma (t,t')
=\frac{1}{2}\int dk~I(k)~ e^{-i k [\bar{r}^{{\star}'}(t)-\bar{r}^{{\star}'}(t')+\bar{t'}(t)-\bar{t'}(t')]}\nonumber\\
&&\;\;\;\;\;\;\;\;\;\;\;\;+\frac{1}{2}\int dk~I(k)~e^{-i k [\bar{r}^{{\star}'}(t')-\bar{r}^{{\star}'}(t)+\bar{t'}(t)-\bar{t'}(t')]}\nonumber\\
&=&\int^{\infty}_0 dk~G(k)\left[\coth\left(\frac{\pi k}{\kappa_{c}}\right) \cos k(t-t')-i \sin k(t-t')\right]~,\nonumber\\
\label{eq450}
\end{eqnarray}
where $G(k)=\frac{2}{\pi \kappa_{c}} \sinh(\pi k/\kappa_{c})\int^{\infty}_0 d k' I(k')~\times$\\$[K_{2 i k'/\kappa_{c}}(2 k' e^{\kappa_{c} \Sigma}/\kappa_{c})+K_{2 i k'/\kappa_{c}}(2 k' e^{-\kappa_{c} \Sigma}/\kappa_{c})]$ (for more details, see Appendix \ref{influencekernel}, with surface gravity $\kappa_{c}$ instead of acceleration $a$ and appropriate coordinate system).
From Eq. (\ref{eq450}), the temperature associated with Hawking radiation for the cosmological horizon comes out to be $T_{c}=\frac{\kappa_{c}}{2 \pi k_{B}}$.
Further, one can use the Unruh vacuum state, as discussed in the previous sections. Same temperatures associated with the black hole and cosmological horizons are obtained, as that using the Hartle-Hawking vacuum sate.

\bigskip
\section{Conclusions}

Generalizing to relativistic
exponential scaling and, using the theory of noise from quantum fluctuations, it was shown that one vacuum (Rindler, Hartle-Hawking,
or Gibbons-Hawking for the cases of the uniformly accelerated detector, black hole, and de Sitter universe, respectively) can be understood as resulting from the scaling of quantum noise in another vacuum.
Here, this idea was explored more generally to establish a flat and curved spacetime analogy. This provides a common perspective to the Unruh and Hawking effects. For this purpose, we examined noise kernels for free fields in some well-known curved spacetimes, e.g., charged black hole, Kerr black hole, Schwarzschild-de Sitter, Schwarzschild anti-de Sitter, and Reissner-Nordstrom de-Sitter spacetimes. We have shown that the exponential scaling transformation is responsible for the thermal nature of radiation. Here, we consider a maximal analytic extension for these spacetimes and different vacuum states. The temperature obtained, exactly matches with the literature in different contexts, using, for example, the Bogoliubov transformations. This formalism can also be applied for some other curved spacetime such as Reissner-Nordstrom-Anti-de Sitter  \cite{Brecher:2004gn, Wang:2004bv}, Kerr-de Sitter  \cite{LI2017211, 10.1143/PTP.100.491}, Kerr-Anti-de Sitter \cite{Das_2000, PhysRevD.98.104013}, Kerr–Newman–de Sitter  \cite{Franzen:2020gke, Gwak:2018tmy,PhysRevD.58.084003}, and Kerr-Newman-Anti-de Sitter spacetimes \cite{Podolsky:2006px, Mangut:2021suk}.  

Recently, it has been shown that quasinormal modes of AdS Black Holes \cite{Horowitz} are related to black hole temperature. It would be interesting to check the relation between the quantum fluctuations to the quasinormal mode for a black hole spacetime.

\begin{acknowledgments}

 CS thanks the Saha Institute of Nuclear Physics (SINP) Kolkata for financial support.
\end{acknowledgments}

\appendix

\section{The exponent of the influence kernel}\label{influencekernel}

The exponent of the influence functional has the following form for the Unruh effect,

\begin{eqnarray}
\varsigma (t(\tau), t(\tau'))
&=&\frac{1}{2}\int dk~I(k)~ e^{-i k [x(\tau)-x(\tau')+t(\tau)-t(\tau')]}\nonumber\\
&&+\frac{1}{2}\int dk~I(k)~ e^{-i k [x(\tau')-x(\tau)+t(\tau)-t(\tau')]}\nonumber\\
&=&\frac{1}{2}\int dk~I(k)~ e^{-i k (U(\tau)-U(\tau'))}
 \nonumber\\
&&+\frac{1}{2}\int dk~I(k)~e^{-i k (V(\tau)-V(\tau'))}\nonumber\\
&=& \frac{1}{2}\int dk~I(k)~ [e^{-2 i k e^{a \Sigma}\sinh(a \Delta)/a}\nonumber\\&&~~~~~~~~~~+e^{-2 i k e^{- a \Sigma}\sinh(a \Delta)/a}]~,
\end{eqnarray}
where  $U=t-x$ and $V=t+x $ are the null coordinates, $2 \Sigma= \tau+\tau'$ and $ \Delta= \tau- \tau'$. Here $U(\tau)-U(\tau')=2 e^{- a \Sigma}\sinh(a \Delta)/a$ and  $V(\tau)-V(\tau')=2 e^{ a \Sigma}\sinh(a \Delta)/a$. Using the expansion 
\begin{eqnarray}
e^{-i \alpha \sinh(x/2)}&=& \frac{4}{\pi}\int^\infty_0 d\nu K_{2 i \nu}(\alpha)[\cosh(\pi \nu) \cos(\nu x)\nonumber\\&&~~~~~~~~~-i \sinh(\pi \nu)\sin(\nu x)]~,\label{Bessel expansion}
\end{eqnarray}
where $K_n$ is the Bessel function of order n,  we get $\varsigma (\tau, \tau')$ as,

\begin{eqnarray}
&&\varsigma (\tau, \tau')\nonumber\\&&=\int^{\infty}_0 dk G(k)[\coth\left(\frac{\pi k}{a}\right) \cos k(\tau-\tau')-i \sin k(\tau-\tau')]~,\nonumber\\
\end{eqnarray} 
where $G(k)=\frac{2}{\pi a} \sinh(\pi k/a)\int^{\infty}_0 d k' I(k')[K_{2 i k'/a}(2 k' e^{a \Sigma}/a) \\+K_{2 i k'/a}(2 k' e^{-a \Sigma}/a)]$.

\section{Unruh vacuum state}\label{unruhsate}
For Unruh vacuum state, one uses null coordinates $U$ and $v$.
The influence kernel for a 1+1 dimensional charged black hole can be written in terms of null coordinates ($U$ and $v$) as,

\begin{eqnarray}\label{unru}
 &&\varsigma (t,t') =\frac{1}{2}\int dk~I(k)~ e^{-i k (U(t)-U(t'))}
 \nonumber\\
&&\;\;\;\;\;\;\;\;\;\;\;\;+\frac{1}{2}\int dk~I(k)~e^{-i k (v(t)-v(t'))}~.
\end{eqnarray}
 For a detector at fixed $r^{\star}$ the influence kernel (\ref{unru}), using the Unruh vacuum state, can be written as,
 
 \begin{eqnarray}
\varsigma (t, t')
&=& \frac{1}{2}\int dk~I(k)~ e^{-2 i k e^{-\kappa_{+}  \Sigma}\sinh(\kappa_{+}\Delta)/\kappa_{+}}\nonumber\\
&&\;\;\;\;\;\;\;\;\;\;\;\;+\frac{1}{2}\int dk~I(k)~e^{-i k \Delta}~,
\end{eqnarray}
where  $2 \Sigma= t+t'$ and $ \Delta= t- t'$.  Here, if we consider $r^{\star}$ zero then $U(t)-U(t')=2 e^{-\kappa_{+}  \Sigma}\sinh(\kappa_{+}\Delta)/\kappa_{+}$ and  $v(t)-v(t')=\Delta$. Using the Eq. (\ref{Bessel expansion}) 
we get $\varsigma (t, t')$ as,
\begin{eqnarray}
&&\varsigma (t,t')\nonumber\\&&=\int^{\infty}_0 dk~G(k)\left[\coth\left(\frac{\pi k}{\kappa_{+}}\right) \cos k(t-t')-i \sin k(t-t')\right]\nonumber\\
&&\;\;\;\;\;\;\;\;\;\;\;\;+\frac{1}{2}\int dk~I(k)~e^{-i k (t-t')}~,
\end{eqnarray}
where $G(k)=\frac{2}{\pi \kappa_{+}} \sinh(\pi k/\kappa_{+})\int^{\infty}_0 d k' I(k')~\times$\\$[K_{2 i k'/\kappa_{+}}(2 k' e^{-\kappa_{+} \Sigma}/\kappa_{+})]$.

\end{document}